\documentclass[reqno]{amsart} 

\usepackage{graphicx}
\usepackage{bm}

\newcommand{\nc}{\newcommand}

\nc{\rnc}{\renewcommand} \nc{\beq}{\begin{equation}}
\nc{\eeq}{\end{equation}} \nc{\beqa}{\begin{eqnarray}}
\nc{\eeqa}{\end{eqnarray}}
\def \z{\underline{z}}
\def \s{\underline{s}}
\def \y{\underline{y}}

\def\stackreb#1#2{\ \mathrel{\mathop{#1}\limits_{#2}}}

\begin{document}

\begin{flushright} AEI-2010-104 \end{flushright}

\title[${\mathcal N}=4$ superconformal indices]
{\bf Superconformal indices of  \\
${\mathcal N}=4$ SYM field theories}

\author{V.~P.~Spiridonov}
\address{Bogoliubov Laboratory of Theoretical Physics,
JINR, Dubna, Moscow Region 141980, Russia and Max-Planck-Institut f\"ur Mathematik,
 Vivatsgasse 7, 53111, Bonn, Germany; e-mail address:
spiridon@theor.jinr.ru}

\author{G.~S.~Vartanov}
\address{Max-Planck-Institut f\"ur Gravitationsphysik, Albert-Einstein-Institut
14476 Golm, Germany; e-mail address: vartanov@aei.mpg.de}

\begin{abstract}
Superconformal indices (SCIs) of $4d$ ${\mathcal N}=4$ SYM theories with simple
gauge groups are described in terms of elliptic hypergeometric integrals.
For $F_4, E_6, E_7, E_8$ gauge groups this yields first examples of integrals
of such type. $S$-duality transformation for $G_2$ and $F_4$ SCIs
is equivalent to a change of integration variables. Equality of SCIs for
$SP(2N)$ and $SO(2N+1)$ group theories is proved in several important
special cases.
Reduction of SCIs to partition functions of $3d$ $\mathcal{N}=2$
SYM theories with one matter field in the adjoint representation
is investigated, corresponding $3d$ dual partners are found,
and some new related hyperbolic beta integrals are conjectured.

\end{abstract}

\maketitle

{Keywords: superconformal index, elliptic hypergeometric integrals,
$S$-duality
\\ \indent 
MSC (2010) codes: 81T60, 33E20}

\section{Introduction}

The problem of electric-magnetic duality for non-abelian gauge theories
was raised by Goddard, Nyuts, and Olive \cite{Goddard} (see also \cite{Montonen}).
Its consideration in the context of $\mathcal{N}=4$
supersymmetric Yang-Mills (SYM) theory in four dimensional
space-time is a quite old area of research \cite{Osborn:1979tq}.
This duality (called also $S$-duality) states the equivalence of the theory
with an ``electric" gauge group $G_c$ to a similar theory with a ``magnetic"
gauge group $G_c^\vee$. Let $G_c$ be a simply laced Lie group.
This means that its Dynkin diagram contains only simple links, and therefore all roots
of the corresponding Lie algebra have the same length, which is true
for $SU(N), SO(2N), E_6, E_7,$ and $E_8$ groups. Then, $G_c^\vee=G_c$
and the $S$-duality transformation maps the complex coupling constant
$\tau =\theta/2\pi + 4\pi\textup{i} /g^2$ to $-1/\tau$. Taken together
with the symmetry transformation $\tau \rightarrow \tau+1,$
the $S$-duality becomes equivalent to the
$SL(2,\mathbb{Z})$-group of modular transformations
\beq \tau \rightarrow
\frac{a\tau+b}{c\tau+d}, \quad ad-bc=1, \ a,b,c,d\in{\mathbb Z}.\eeq

For the non-simply laced gauge groups, the $S$-duality acts as
$\tau \rightarrow -1/m\tau$, where $m$ is the ratio of
the lengths-squared of long and short roots of the corresponding
root system. One has $m=2$ for $SO(2N+1)$ and $SP(2N)$ group theories
dual to each other \cite{Goddard}. For $F_4$ and $G_2$ groups one has
$m=2$ and $m=3$, respectively; corresponding theories were discussed
in \cite{AKS} from the algebraic point of view and the $S$-duality
transformation of their moduli spaces was described.

Here we discuss a new test of $\mathcal{N}=4$ SYM field theory dualities
based on the superconformal indices (SCIs) suggested by
Kinney et al in \cite{Kinney} (for the definition of indices in
$\mathcal{N}=1$ theories, see \cite{Romelsberger1}).
$\mathcal{N}=4$ SYM theory has the $PSU(2,2|4)$ space-time symmetry
group generated by $J_a, \overline{J}_a$, $a=1,2,3$, representing
$SU(2)$ subgroups (Lorentz rotations), $P_\mu,
Q_{i,\alpha},\overline{Q}_{i,\dot\alpha}$ (supertranslations) with
$\mu=0,1,2,3,\ i=1,2,3,4$ and $\alpha,\dot\alpha=1,2$, $K_\mu,
S_{i,\alpha},\overline{S}_{i,\dot\alpha}$ (special superconformal
transformations), and $H$ (dilations) whose  state eigenvalues are given
by conformal dimensions \cite{Dolan:2002zh}. As to the $SU(4)_R$
$R$-symmetry subgroup, we mention only its commuting maximal torus
generators $R_1,R_2,R_3$. For a distinguished pair of supercharges,
say, $Q:= Q_{1,1}$ and $Q^{\dag}:=S_{1,1}$, in appropriate
normalization one has
\begin{equation}
\{Q,Q^{\dag}\}= H - 2J_3 - 2\sum_{k=1}^3
\left(1-\frac{k}{4}\right)R_k =: \Delta.
\label{susy}\end{equation}
In this case SCI is defined as the following gauge-invariant trace
 \beq\label{SI}
I(t,y,v,w) = \text{Tr} \left( (-1)^{\mathcal F} t^{2(H+J_3)}
y^{2\overline{J}_3} v^{R_2} w^{R_3} e^{-\beta\Delta} \right),\eeq
where ${\mathcal F}$ is the fermion number
operator and $t,y,v,w, g_a, \beta$ are group parameters (chemical potentials).
The trace is effectively taken over the space of zero modes of the operator
$\Delta$ (the space of BPS states \cite{D}), because relation (\ref{susy})
is preserved by operators used in (\ref{SI}); the contributions from other
states cancel together with the dependence on $\beta$.
In comparison to $\mathcal{N}=1, 2$ theories, all fields of $\mathcal{N}=4$
SYM theory lie in the adjoint representation of $G_c$, i.e. only the adjoint
representation characters enter SCIs.

The $U(N)$-gauge group SCI has the following matrix integral form \cite{Kinney}
\beq \label{Ind}
I(t,y,v,w) = \int_{G_c} [dU]\,  \exp \Big( \sum_{m=1}^\infty \frac 1m
f(t^m,y^m,v^m,w^m) \text{Tr}(U^\dag)^m \text{Tr}\,  U^m\Big),\eeq where $[dU]$ is
the invariant measure and $f(t,y,v,w)\text{Tr}\,  U^\dag \text{Tr}\,  U$ is the so-called
single-particle states index with
$$
f(t,y,v,w) = \frac{t^2(v+1/w + w/v) - t^3 (y+1/y)
- t^4 (w+1/v+v/w) + 2 t^6}{(1-t^3y)(1-t^3/y)}.
$$
As shown in \cite{Bianchi:2006ti} (see there the discussion following formula
(5.33)), this expression can be obtained from the superconformal group
character or partition function for $\mathcal{N}=4$ theories by imposing
the shortening condition for the multiplets.

The SCI technique has found many
applications in supersymmetric field theories. R\"omelsberger
conjectured \cite{Romelsberger1} that SCIs of the Seiberg dual $\mathcal{N}=1$ SYM
theories coincide. Dolan and Osborn explicitly confirmed this
conjecture for a number of examples  \cite{Dolan:2008qi}. It appeared
that SCIs are expressed in terms of elliptic
hypergeometric integrals whose theory was developed earlier in
\cite{S1,S2} (see also \cite{S3} for a general survey).
Equality of indices in dual theories happened to be equivalent
either to exact computability of elliptic beta integrals discovered
in \cite{S1} or to nontrivial Weyl group symmetry transformations
for higher order elliptic hypergeometric functions \cite{S2,Rains}.
In a series of papers \cite{SV2,SV} we applied this technique
to analyzing all previously found Seiberg dualities. We suggested
also many new such dualities  on the basis of known identities for
elliptic hypergeometric integrals and showed that known
nontrivial duality checks are satisfied for them.
As a payback to mathematics, it happened that many old dualities lead
to new, still unproven highly nontrivial relations for integrals.

This line of thoughts was further developed in beautiful papers by
Gadde et al \cite{GPRR1,GPRR2}. In \cite{Bult}, a particular
one dimensional elliptic hypergeometric integral was shown to have
$W(F_4)$ Weyl group of symmetry, which follows from the
elliptic beta integral evaluation formula \cite{S1}. It was used in \cite{GPRR1}
for confirming $S$-duality for $\mathcal{N}=2$ SYM theory with
$SU(2)$ gauge group and four hypermultiplets and for ensuring associativity
of the operator algebra of $2d$ topological field theories behind that duality.
The SCI for a $E_6$ SCFT theory was constructed in \cite{GPRR2}
from the index of $\mathcal{N}=2$ SYM theory with $G_c=SU(3)$ and six hypermultiplets
and a new test of the Argyres-Seiberg duality was suggested.

Here we construct $\mathcal{N}=4$ SCIs
for all simple gauge groups, show their $S$-duality invariance for $G_2$ and $F_4$
cases, and give new mathematical arguments
supporting equality of  SCIs for $SP(2N)$ and $SO(2N+1)$ theories
conjectured in \cite{GPRR1}.
All $\mathcal{N}=4$ indices degenerate in a specific limit
to orthogonality measures for the Macdonald polynomials and admit
thus exact evaluations. Another limit leads to computable $3d$ partition functions
described by the hyperbolic beta integrals.

\section{Duality of $SO(2N+1)$ and $SP(2N)$  $\mathcal{N}=4$ SYM theories}

SCIs for $SP(2N)$ and $SO(2N+1)$ $\mathcal{N}=4$ SYM theories were described
in \cite{GPRR1} and discussed briefly in the simplest case in
\cite{SV2}. Here we prove equality of these SCIs in several important
limiting cases.

In all $\mathcal{N}=4$ theories the single-particle index is
\beq
\frac{1}{(1-p)(1-q)}\Big(\sum_{k=1}^3 s_k - pq\sum_{k=1}^3 s_k^{-1}
-p-q + 2 pq\Big)\chi_{adj}(z),\eeq
where $\chi_{adj}(z)$ is the character of the adjoint representation of
the corresponding gauge group (see the Appendix). For convenience,
we have denoted
$$
s_1 =t^2 v, \quad s_2 =t^2w^{-1}, \quad
s_3=t^2 wv^{-1}, \quad p =t^3y, \quad q=t^3y^{-1}.
$$
Using explicit expressions of the group invariant measures, SCIs can be written
as particular elliptic hypergeometric integrals \cite{S3}.
So, $SP(2N)$-electric theory index gets the following form
\beq
\label{SOSP_1} I_E = \chi_N \int_{\mathbb{T}^N} \prod_{1 \leq i < j
\leq N} \frac{\prod_{k=1}^3
\Gamma(s_kz_i^{\pm1}z_j^{\pm1};p,q)}{\Gamma(z_i^{\pm1}z_j^{\pm1};p,q)}
\prod_{j=1}^N \frac{\prod_{k=1}^3
\Gamma(s_kz_j^{\pm2};p,q)}{\Gamma(z_j^{\pm2};p,q)}
\frac{dz_j}{2 \pi \textup{i} z_j}, \eeq
and for $SO(2N+1)$-magnetic theory one has
\beq
\label{SOSP_2} I_M = \chi_N \int_{\mathbb{T}^N} \prod_{1 \leq i < j
\leq N} \frac{\prod_{k=1}^3 \Gamma(s_k
y_i^{\pm1}y_j^{\pm1};p,q)}{\Gamma(y_i^{\pm1}y_j^{\pm1};p,q)}
\prod_{j=1}^N \frac{\prod_{k=1}^3 \Gamma(s_k
y_j^{\pm1};p,q)}{\Gamma(y_j^{\pm1};p,q)} \frac{dy_j}{2
\pi \textup{i} y_j},\eeq
where  $|s_k|<1,\, k=1,2,3.$ For $|s_k|\geq 1$ the indices are defined as
analytical continuations of the expressions \eqref{SOSP_1} and \eqref{SOSP_2}.
Here $\mathbb{T}$ denotes the unit circle with positive orientation
and we use conventions
$\Gamma(a,b;p,q):=\Gamma(a;p,q)\Gamma(b;p,q),$
$\Gamma(az^{\pm1};p,q):=\Gamma(az;p,q)\Gamma(az^{-1};p,q)$,
where
$$
\Gamma(z;p,q)= \prod_{i,j=0}^\infty
\frac{1-z^{-1}p^{i+1}q^{j+1}}{1-zp^iq^j}, \quad |p|, |q|<1,
$$
is the elliptic gamma function. The coefficient in front of the
integrals is
$$\chi_N \ = \ \frac{(p;p)^N_\infty (q;q)^N_\infty}{2^N
N!} \prod_{k=1}^3 \Gamma^{N}(s_k;p,q),$$  with
$(a;q)_\infty=\prod_{k=0}^\infty(1-aq^k)$.
The constraint $\prod_{k=1}^3 s_k = pq$
plays the role of the balancing condition for integrals.

The $S$-duality hypothesis leads thus to the conjecture $I_E = I_M$, or
\begin{eqnarray} &&
\int_{\mathbb{T}^{N}} \Delta_E(\z,\s) \prod_{j=1}^{N} \frac{d z_j}{2
\pi \textup{i} z_j} = \int_{\mathbb{T}^N} \Delta_M(\y,\s) \prod_{j=1}^{N}
\frac{d y_j}{2 \pi \textup{i} y_j},
\end{eqnarray}
where the kernels $\Delta_E(\z,\s)$ and $\Delta_M(\y,\s)$ are read
from integrals (\ref{SOSP_1}) and (\ref{SOSP_2}).
Denoting $\rho(\z,\y,\s) =\Delta_E(\z,\s)/\Delta_M(\y,\s)$,
we have verified that this function represents the so-called totally elliptic
hypergeometric term \cite{SMar,SV2}. This is a rather rich
mathematical statement giving strong evidence on the validity of
the stated equality of integrals. It means that all the functions
\begin{eqnarray*}
&& h_i^{(z)}=\frac{\rho(\dots qz_i \dots,\y,\s)}{\rho(\z,\y,\s)}, \quad
h_i^{(y)}=\frac{\rho(\z,\dots qy_i \dots,\s)}{\rho(\z,\y,\s)}, \quad
i=1,\ldots,N,
\\ \makebox[0em]{}
&& h^{(s)}_{kl}=\frac{\rho(\z,\y,\ldots qs_k, \ldots, q^{-1}s_l
\ldots)}{\rho(\z,\y,\s)}, \quad k,l=1,2,3, \ k \neq l,
\end{eqnarray*}
are elliptic functions of all their arguments $z_i, y_i,s_k,$ and
$q$. For instance,
\begin{eqnarray*}
&&
h_i^{(z)}(\z,\y,\s;q;p)=h_i^{(z)}(\dots pz_j\dots,\y,\s;q;p)
= h_i^{(z)}(\z,\dots  py_j\dots,\s;q;p)
\\ && \makebox[2em]{}
 =h_i^{(z)}(\z,\y,\ldots ps_k \ldots p^{-1}s_l;q;p)
= h_i^{(z)}(\z,\y,\ldots ps_l \ldots;pq;p),
\end{eqnarray*}
where $k,l=1,2,3$.
This test is passed by all known integral identities, though it is not
sufficient for their validity.
For further consequences of the total ellipticity and various
technical details of such computations, see \cite{S3,SV2,SMar}.

Now we list various special cases when the equality $I_E=I_M$ can be verified
rigorously. For low ranks of the gauge group, it follows from the change of
integration variables
associated with the affine transformation of the corresponding root system \cite{GPRR1}.
The electric SCI is obtained from the magnetic one after the substitution $y=z^2$
for $N=1$, and $y_1=z_1z_2$ and $y_2 =z_1/z_2$ for $N=2$.

{\bf The limit $s_k\to1$.} Suppose that one of the parameters, say, $s_1$ goes to $1$.
Then elliptic gamma functions of the integrand denominators
are cancelled and no singularities appear on the integration contour.
Because now $s_2s_3=pq$, and $\Gamma(a,b;p,q)=1$ for $ab=pq$,
the integrands are actually equal to 1. However, the factor $\chi_N$
is divergent in this limit.  As a result, we have $\lim_{s_1\to
1}I_E/I_M=1$.

{\bf Reduction $p=q=0$.} Consider the limit $p\to 0$.
For fixed $z$, the limit $p\to0$ and further limit $q\to0$ simplifies
the elliptic gamma function to
$$
 \Gamma(z;p,q) \stackreb{=}{p\to
0} \frac{1}{(z;q)_\infty}  \stackreb{=}{q\to 0} \frac{1}{1-z}.
$$
Because of the balancing condition for integrals, all parameters
cannot be kept fixed. The simplest possibility consists in fixing
$s_{1,2}$ and setting
$s_3 = pq/s_1s_2.$ Then integral (\ref{SOSP_1}) reduces to
\beqa
\label{SOSPq1} && \makebox[-2em]{}
 I_E^{p=0} (s_1, s_2 \mbox{ fixed}) =
\frac{(q;q)_\infty^N}{2^N N!} \frac{(s_1s_2;q)_\infty^N}{(s_1,s_2;q)_\infty^N}
\\ \nonumber &&
\times \int_{\mathbb{T}^N} \prod_{1 \leq i < j
\leq N} \frac{(z_i^{\pm1}z_j^{\pm1},s_1s_2z_i^{\pm1}z_j^{\pm1};q)_\infty}
{(s_1z_i^{\pm1}z_j^{\pm1},s_2z_i^{\pm1}z_j^{\pm1};q)_\infty}
\prod_{j=1}^N \frac{(z_j^{\pm2},s_1s_2z_j^{\pm2};q)_\infty}
{(s_1z_j^{\pm2},s_2z_j^{\pm2};q)_\infty} \frac{dz_j}{2 \pi \textup{i} z_j},
\eeqa
where $(a,b;q)_\infty:=(a;q)_\infty(b;q)_\infty$.
Integral (\ref{SOSP_2}) reduces to
\beqa \label{SOSPq2} && \makebox[-2em]{}
I_M^{p=0} (s_1,
s_2 \mbox{ fixed}) = \frac{(q;q)_\infty^N}{2^N N!}
\frac{(s_1s_2;q)_\infty^N}{(s_1,s_2;q)_\infty^N}  \\ \nonumber
&&
\times \int_{\mathbb{T}^N} \prod_{1
\leq i < j \leq N} \frac{(y_i^{\pm1}y_j^{\pm1},s_1s_2y_i^{\pm1}y_j^{\pm1};q)_\infty}
{(s_1y_i^{\pm1}y_j^{\pm1},s_2y_i^{\pm1}y_j^{\pm1};q)_\infty}
\prod_{j=1}^N \frac{(y_j^{\pm1},s_1s_2y_j^{\pm1};q)_\infty}
{(s_1y_j^{\pm1},s_2y_j^{\pm1};q)_\infty} \frac{dy_j}{2 \pi \textup{i} y_j}.
\eeqa

For $q=0$ the integrands have only a finite number of poles and
the integrals can be evaluated by computing the residues. However, we did
not find a simple way of performing these computations for arbitrary $N$
and have verified equality of the resulting $p=q=0$ SCIs only for $N=3$.

One can tie the limit $p,q\to0$ to a very
natural choice of the fugacities $v, w$ in (\ref{SI}) equal to $1$.
After fixing $s_k = (pq)^{1/3}, k=1,2,3,$ the limit $p, q\to 0$ strongly
simplifies the integrals (set $q=s_1=s_2=0$ in \eqref{SOSPq1} and \eqref{SOSPq2}).
Then the SCIs can be evaluated exactly using two different
special cases of the Selberg integral, description of
which we skip for brevity, yielding $I_E=I_M=1$.

{\bf A  $p=0,\, q \to 1$ limit.} Let us set in \eqref{SOSPq1}, \eqref{SOSPq2}
$s_1=q^\alpha,\, s_2=q^{\beta}$ and consider the limit $q\to 1$
for fixed $\alpha$ and $\beta$. Known asymptotic formulas
$$
\lim_{q \rightarrow 1} \frac{(q^\alpha z;q)_{\infty}}{(q^{\beta}z;q)_{\infty}} =
(1-z)^{\beta-\alpha}, \qquad
\lim_{q
\rightarrow 1}\frac{(q;q)_{\infty}}{(q^{x};q)_{\infty}}(1-q)^{1-x} =
\Gamma(x),
$$
where $\Gamma(x)$ is the Euler gamma function,
show that both integrands become equal to 1 and the leading asymptotics
for SCIs is determined by the integral prefactors
$$
I^{p=0,q \rightarrow 1}_{E,M} (s_1 = q^{\alpha}, s_2 =
q^{\beta}) = \frac{1}{2^NN!}\left(\frac{\Gamma(\alpha)\Gamma(\beta)}
{(1-q)\Gamma(\alpha+\beta)}\right)^N(1+ o(1)).
$$

{\bf A  $p=0,\, s_2=0$ limit.} Let us set now in \eqref{SOSPq1} $s_2=0$, which yields
\beq \label{SP_lim3d}\makebox[-2em]{}
I_{SP(2N)}^{p=s_2=0} =
\frac{1}{2^{N}N!}
\frac{(q;q)^{N}_\infty}{(s_1;q)_\infty^{N}}
\int_{\mathbb{T}^{N}} \prod_{1 \leq i < j \leq N}
\frac{(z_i^{\pm1}z_j^{\pm1};q)_\infty}{(s_1z_i^{\pm1}z_j^{\pm1};q)_\infty}
\prod_{j=1}^N \frac{(z_j^{\pm2};q)_\infty}{(s_1z_j^{\pm2};q)_\infty}
 \frac{dz_j}{2 \pi \textup{i} z_j}. \eeq
This integral can be evaluated exactly using the multivariable extension of
the Askey-Wilson integral (or particular $q$-Selberg integral
serving as the orthogonality measure for Koornwinder polynomials) found
in \cite{Gustafson}
\beqa\label{gus} && \makebox[-2em]{} \frac{1}{2^N N!}
\int_{\mathbb{T}^N} \prod_{1 \leq i < j \leq N}
\frac{(z_i^{\pm1}z_j^{\pm1};q)_{\infty}}{(bz_i^{\pm1}z_j^{\pm1};q)_{\infty}}
\prod_{j=1}^N \frac{(z_j^{\pm2};q)_{\infty}}{\prod_{i=1}^4
(a_iz_j^{\pm1};q)_{\infty}} \frac{dz_j}{2\pi \textup{i} z_j}
\nonumber \\ &&
= \prod_{j=1}^N \Big(
\frac{(t;q)_{\infty}(b^{N+j-2}a_1a_2a_3a_4;q)_{\infty}}
{(b^j;q)_{\infty}(q;q)_{\infty}}  \prod_{1 \leq i < k \leq 4}
\frac{1}{(b^{j-1}a_ia_k;q)_{\infty}} \Big),\eeqa
where $|b|, |a_i|<1$.
This formula reduces to our case after the substitutions
$$
b=s_1,\qquad a_{1,2}=\pm \sqrt{s_1},\qquad  a_{3,4}=\pm \sqrt{qs_1}.
$$

The same limit applied to \eqref{SOSPq2} leads to the integral
\beq \label{SO_lim3d}\makebox[-2em]{}
I_{SO(2N+1)}^{p=s_2=0} = \frac{1}{2^{N}N!}
\frac{(q;q)^{N}_\infty}{(s_1;q)_\infty^{N}}
 \int_{\mathbb{T}^{N}} \prod_{1 \leq i < j \leq N}
\frac{(z_i^{\pm1}z_j^{\pm1};q)_\infty}{(s_1z_i^{\pm1}z_j^{\pm1};q)_\infty}
\prod_{j=1}^N \frac{(z_j^{\pm1};q)_\infty}{(s_1z_j^{\pm1};q)_\infty}
\frac{dz_j}{2 \pi \textup{i} z_j}, \eeq
which is obtained from \eqref{gus} after setting
$$
b=s_1,\qquad a_1=s_1,\qquad a_2=-1,\qquad  a_{3,4}=\pm \sqrt{q}.
$$
Corresponding computations on the right-hand side of \eqref{SOSPq2} yield
\beq
I_{SP(2N)}^{p= s_2=0} \ = \   I_{SO(2N+1)}^{p=s_2=0} \ = \ \prod_{j=0}^{N-1}
\frac{(qs_1^{2j+1};q)_\infty}{(s_1^{2j+2};q)_\infty}.
\label{spso}\eeq
Equality of indices established earlier in the limit $s_k=(pq)^{\frac 13}\to 0$, $k=1,2,3$,
is a special case of relation \eqref{spso} obtained after fixing $s_1=q=0$.

 The integrals in \eqref{spso} were computed under
the assumption that $|s_1|<1$, but for finite $N$ we can analytically continue
SCIs to arbitrary values of $s_1$
 as meromorphic functions using the right-hand side
expression. For $|s_1|<1$, the limit $N\to\infty$
yields a ratio of double infinite products appearing in the elliptic gamma function
with $p=s_1^2$. From the physical point of view this limit is
relevant for testing the AdS/CFT correspondence. In \cite{Nakayama}, it was
suggested to consider the maximal angular momentum limit for indices
$t \rightarrow 0,$ $y \rightarrow \infty$ with $t^3y$ fixed, which
corresponds to $q\to 0$ with fixed $p$. Due to the symmetry between $p$ and $q$
this is similar to our limit $p=s_2=0$, but we have the additional free
parameter $s_1$ absent in \cite{Nakayama}.

{\bf The hyperbolic limit.} Let us study the hyperbolic limit \cite{ds,rai:limits}
of elliptic hypergeometric integrals (\ref{SOSP_1}) and (\ref{SOSP_2}).
First we parametrize the variables as
$$
p = e^{2 \pi \textup{i} v \omega_1}, \quad q = e^{2 \pi \textup{i} v \omega_2},
\quad  s_i = e^{2 \pi \textup{i} v \alpha_i}, \; i=1,2,3,
$$
where $\sum_{i=1}^3 \alpha_i = \omega_1 + \omega_2$ (the balancing condition),
and then take the limit $v \rightarrow 0$. To simplify the integrals we
use the Ruijsenaars limit
\beq \Gamma(e^{2 \pi \textup{i} r z};e^{2 \pi \textup{i} r \omega_1},
e^{2 \pi \textup{i} r \omega_2}) \stackreb{=}{r \rightarrow 0}
e^{-\pi \textup{i}(2z-\omega_1-\omega_2)/12r\omega_1\omega_2}
\gamma^{(2)}(z;\omega_1,\omega_2),
\eeq
where
\beq
 \gamma^{(2)}(u;\omega_1,\omega_2) = e^{-\frac{\pi \textup{i}}{2}
B_{2,2}(u;\omega_1,\omega_2)} \frac{(e^{2 \pi \textup{i}(u-\omega_2)/\omega_1};
e^{-2 \pi \textup{i} \omega_2/\omega_1})_\infty}{(e^{2 \pi \textup{i}
u/\omega_2};e^{2 \pi \textup{i} \omega_1/\omega_2})_\infty}\eeq
is the hyperbolic gamma function and $B_{2,2}(u;\mathbf{\omega})$
is the second order Bernoulli polynomial,
$$
 B_{2,2}(u;\mathbf{\omega}) =
\frac{u^2}{\omega_1\omega_2} - \frac{u}{\omega_1} -
\frac{u}{\omega_2} + \frac{\omega_1}{6\omega_2} +
\frac{\omega_2}{6\omega_1} + \frac 12.
$$
The following conventions are used below
$
\gamma^{(2)}(a,b;\mathbf{\omega}) :=
\gamma^{(2)}(a;\mathbf{\omega}) \gamma^{(2)}(b;\mathbf{\omega})
$
and
$
\gamma^{(2)}(a\pm u;\mathbf{\omega}) :=
\gamma^{(2)}(a+u;\mathbf{\omega})
\gamma^{(2)}(a-u;\mathbf{\omega}).
$

We skip the general expressions for hyperbolic integrals arising
in this limit
and present only the result appearing after taking the additional limit
$\alpha_2 \rightarrow \infty$ (which mimics altogether the previously considered
limit $p=0, s_2 = 0$):
\beq \label{hyperSP}
I_{SP(2N)}^{h,\alpha_2 \rightarrow \infty} =
 \xi_N \int_{- \textup{i} \infty}^{\textup{i}\infty}
\prod_{1 \leq i < j \leq N} \frac{\gamma^{(2)}( \alpha_1 \pm u_i \pm u_j;\mathbf{\omega})}
{\gamma^{(2)}( \pm u_i \pm u_j;\mathbf{\omega})} \prod_{j=1}^N
\frac{\gamma^{(2)}( \alpha_1 \pm 2 u_j;\mathbf{\omega})}
{\gamma^{(2)}( \pm 2 u_j;\mathbf{\omega})} du_j,
\eeq
\beq \label{hyperSO}
I_{SO(2N+1)}^{h,\alpha_2 \rightarrow \infty}
= \xi_N \int_{- \textup{i} \infty}^{\textup{i} \infty} \prod_{1 \leq i < j \leq N}
 \frac{\gamma^{(2)}( \alpha_1 \pm u_i \pm u_j;\mathbf{\omega})}
{\gamma^{(2)}( \pm u_i \pm u_j;\mathbf{\omega})} \prod_{j=1}^N
\frac{\gamma^{(2)}( \alpha_1 \pm u_j;\mathbf{\omega})}
{\gamma^{(2)}( \pm u_j;\mathbf{\omega})} du_j,
\eeq
where $\xi_N = \gamma^{(2)}(\alpha_1;\mathbf{\omega})^N/N!(2 \textup{i}
\sqrt{\omega_1\omega_2})^N$ and we dropped the common multiplier
$
\exp\{\frac{\pi \textup{i}}{2}( \alpha_1^2 + 2 \alpha_1 \alpha_2
- \alpha_1 (\omega_1+\omega_2))(2N^2+N)\}.
$
To obtain these expressions we used the inversion
relation $\gamma^{(2)}(z,\omega_1+\omega_2-z;\mathbf{\omega}) = 1$ and the
asymptotic formulas
\beqa
\makebox[-1.25em]{} \lim_{u \rightarrow \infty}
e^{\frac{\pi \textup{i}}{2} B_{2,2}(u;\mathbf{\omega})} \gamma^{(2)}(u;\mathbf{\omega})
& = & 1, \text{ \ \ for } \text{arg }\omega_1 < \text{arg } u < \text{arg }\omega_2 + \pi, \nonumber \\
\lim_{u \rightarrow \infty}e^{-\frac{\pi \textup{i}}{2} B_{2,2}(u;\mathbf{\omega})} \gamma^{(2)}(u;\mathbf{\omega})
& = & 1, \text{  \ \ for } \text{arg } \omega_1 - \pi < \text{arg } u <
\text{arg }\omega_2.
\eeqa

The following hyperbolic analog of the Selberg integral was computed in \cite{ds}
(for $N=1$, see \cite{rui:int}):
\beqa  && \makebox[-2em]{}
\frac{1}{2^N N!} \int_{-\textup{i}
\infty}^{\textup{i} \infty} \prod_{1 \leq i < k \leq N}
\frac{\gamma^{(2)}(\tau \pm u_i \pm
u_k;\mathbf{\omega})}{\gamma^{(2)}(\pm u_i \pm
u_k;\mathbf{\omega})}
\prod_{j=1}^N \frac{\prod_{i=1}^{4}
\gamma^{(2)}(\mu_i \pm u_j;\mathbf{\omega})}{\gamma^{(2)}(\pm 2
u_j;\mathbf{\omega})} \frac{d u_j}{\textup{i}
\sqrt{\omega_1 \omega_2}} \nonumber \\ && \makebox[2em]{}
= \prod_{j=1}^{N} \frac{\gamma^{(2)}(j \tau;\mathbf{\omega})}
{\gamma^{(2)}(\tau;\mathbf{\omega})}
 \prod_{j=0}^{N-1} \frac{\prod_{1 \leq i < k \leq 4}
\gamma^{(2)}(j \tau + \mu_i + \mu_k;\mathbf{\omega})}{\gamma^{(2)}((2N-2-j)\tau
+ \sum_{i=1}^4 \mu_i;\mathbf{\omega})},
\label{conf2} \eeqa
where the Mellin-Barnes integration contour separates sequences of integrand poles
going to infinity.
One can obtain integral (\ref{hyperSP}) from (\ref{conf2}) after the substitutions
$$
\tau = \alpha_1, \ \mu_1 =\frac 12 \alpha_1, \ \mu_2 = \frac 12(\alpha_1 +
\omega_1), \ \mu_3 = \frac 12 (\alpha_1+ \omega_2), \ \mu_4 =
\frac 12 (\alpha_1 + \omega_1+\omega_2),
$$
and integral (\ref{hyperSO}) after the substitutions
$$
\tau = \alpha_1, \quad \mu_1 = \alpha_1, \quad \mu_2 = \frac 12 \omega_1,
\quad \mu_3 = \frac 12 \omega_2, \quad \mu_4 = \frac 12 (\omega_1+\omega_2)
$$
and application of the duplication formula
$\gamma^{(2)}(2z;\mathbf{\omega}) =
\gamma^{(2)} (z,z + \omega_1/2, z + \omega_2/2, z
+ (\omega_1+\omega_2)/2;\mathbf{\omega})$.
Direct computations show that
\beq \label{hyperbolLim}
I_{SP(2N)}^{h,\alpha_2 \rightarrow \infty} =
I_{SO(2N+1)}^{h,\alpha_2 \rightarrow \infty}
= \prod_{j=0}^{N-1} \frac{\gamma^{(2)}( (2j+2) \alpha_1;\mathbf{\omega})}
{\gamma^{(2)}( (2j+1) \alpha_1 + \omega_1 + \omega_2;\mathbf{\omega})}.
\eeq
Relations \eqref{spso} and \eqref{hyperbolLim} provide the best
available SCI justifications of the duality of $\mathcal{N}=4$ SYM field
theories with $SP(2N)$ and $SO(2N+1)$ gauge groups.

Discuss now a physical interpretation of integrals (\ref{hyperSP}), (\ref{hyperSO})
and their exact evaluation (\ref{hyperbolLim}).
In \cite{DSV} it was shown that the hyperbolic limit of $4d$ $\mathcal{N}=1$
SCIs leads to partitions functions of $3d$ $\mathcal{N}=2$
SYM and CS theories constructed in \cite{Jafferis,Hama} following \cite{KWY}.
Our hyperbolic integrals describe partition functions of $3d$ $\mathcal{N}=2$
SYM theories with $SP(2N)$ and $SO(2N+1)$ gauge groups containing one chiral
superfield in the adjoint representation with the
$U(1)_A$-group hypercharge 1. First, these $3d$
theories are dual to each other and, second, they share the same confining phase
described by a Wess-Zumino type model with $2N$ chiral fields
with the $U(1)_A$-hypercharges $2k, -2k+1$, $k=1,\ldots,N,$ and zero $R$-charges,
whose partition function is given by expression (\ref{hyperbolLim}). Taking
$\alpha_1=(\omega_1+\omega_2)/2$ in (\ref{hyperSP}) and (\ref{hyperSO})
one obtains partition functions for pure $3d$ $\mathcal{N}=4$ SYM
theories. As follows from the exact evaluation (\ref{hyperbolLim}),
these partition functions vanish indicating thus to the spontaneous
supersymmetry breaking \cite{MN}.

As to the hyperbolic integrals obtained from SCIs for arbitrary
$\alpha_{1}$ and $\alpha_2$, they describe partition functions of
$3d$ $\mathcal{N}=2$ SYM theories with 3 chiral superfields in the adjoint
representation. The constraint $\alpha_1=(\omega_1+\omega_2)/2$
leads to partition functions of $3d$ $\mathcal{N}=4$ SYM theories with one
hypermultiplet in the adjoint representation. In these cases, $3d$ theories with
$SP(2N)$ and $SO(2N+1)$ gauge group are dual to each other in the same way as
the parent $4d$ $\mathcal{N}=4$ models. A similar situation holds for all other
cases considered below.

\section{$G_2$ gauge group}

We consider now the $S$-duality conjecture for $\mathcal{N}=4$
SYM theory with the gauge group $G_2$. This group has two maximal torus
variables $z_1$ and $z_2$, but it is convenient to introduce the third
variable $z_3=z_1^{-1}z_2^{-1}$ (see the Appendix). Then
the electric SCI takes the form
\beq
\label{G2_1} I_E = \kappa_2 \int_{\mathbb{T}^2} \prod_{1 \leq i < j
\leq 3} \frac{\prod_{k=1}^3 \Gamma(s_k
z_i^{\pm1}z_j^{\pm1};p,q)}{\Gamma(z_i^{\pm1}z_j^{\pm1};p,q)}
\prod_{j=1}^2 \frac{dz_j}{2 \pi \textup{i} z_j}, \eeq
where $|s_k|<1,\, k=1,2,3,$ and
$$
\kappa_2 \ = \
\frac{(p;p)^2_\infty (q;q)^2_\infty}{2^2 3} \prod_{k=1}^3
\Gamma^2(s_k;p,q).$$
In the magnetic theory one has
\beq \label{G2_2} I_M \ = \ \kappa_2
\int_{\mathbb{T}^2} \prod_{1 \leq i < j \leq 3} \frac{\prod_{k=1}^3
\Gamma(s_k (y_iy_j)^{\pm3}, s_k (y_iy_j^{-1})^{\pm1};p,q)}{\Gamma(
(y_iy_j)^{\pm3},(y_iy_j^{-1})^{\pm1};p,q)} \prod_{j=1}^2
\frac{dy_j}{2 \pi \textup{i} y_j},\eeq
where $y_1y_2y_3= 1$ (we are indebted to S. Razamat for pointing
to a misprint in our initial expression for this integral).

The $S$-duality hypothesis assumes the equality of these elliptic
hypergeometric integrals, $ I_E = I_M$. Remarkably, this identity can be
easily established by the following change of the integration variables
$$
y_1 \  = \ (z_2z_3^2)^{1/3},\qquad y_2 \ = \ (z_3z_1^2)^{1/3},
\qquad  y_3 \ = \ (z_1z_2^2)^{1/3},
$$
associated with the rotation of the $G_2$ root system
\cite{AKS}. The SCI test confirms thus the $S$-duality in this case.

Application of the limit $p=s_2=0$ reduces integral (\ref{G2_1}) to
\beq
\label{G2_1_lim3d} I_{G_2}^{p=s_2=0} \ = \  \frac{1}{2^2 3}
\frac{(q;q)^2_\infty}{(s_1;q)_\infty^2}
\int_{\mathbb{T}^2} \prod_{1 \leq i < j \leq 3}
\frac{(z_i^{\pm1}z_j^{\pm1};q)_\infty}{(s_1
z_i^{\pm1}z_j^{\pm1};q)_\infty} \prod_{j=1}^2 \frac{dz_j}{2 \pi
\textup{i} z_j}, \eeq
 where $z_1z_2z_3=1$. This integral admits exact evaluation \cite{Ito}
\beqa
I_{G_2}^{p=s_2=0} \ = \
\frac{(qs_1,qs_1^5;q)_\infty}{(s_1^2,s_1^6;q)_\infty}.\eeqa

\section{$F_4$ gauge group}

Consider the $S$-duality for $\mathcal{N}=4$ SYM theory with the
gauge group $F_4$ \cite{Goddard,Montonen,Osborn:1979tq,AKS}.
The electric SCI has the following form
\beqa
\label{F4} && I_E \ = \ \kappa_4 \int_{\mathbb{T}^4} \prod_{1 \leq i
< j \leq 4} \frac{\prod_{k=1}^3 \Gamma(s_k
z_i^{\pm2}z_j^{\pm2};p,q)}{\Gamma(z_i^{\pm2}z_j^{\pm2};p,q)}
\prod_{j=1}^4
\frac{\prod_{k=1}^3 \Gamma(s_k z_j^{\pm2};p,q)}{\Gamma(z_j^{\pm2};p,q)} \nonumber \\
&& \makebox[5em]{} \times \frac{\prod_{k=1}^3 \Gamma(s_k
z_1^{\pm1}z_2^{\pm1}z_3^{\pm1}z_4^{\pm1};p,q)}
{\Gamma(z_1^{\pm1}z_2^{\pm1}z_3^{\pm1}z_4^{\pm1};p,q)}
\prod_{j=1}^4 \frac{dz_j}{2 \pi \textup{i} z_j}, \eeqa
 where $|s_k|<1,\, k=1,2,3,$ and
$$
\kappa_4 =
\frac{(p;p)^4_\infty (q;q)^4_\infty}{2^{7} 3^2} \prod_{k=1}^3\Gamma^{4} (s_k;p,q).
$$
In the derivation of this expression we used the $F_4$ group adjoint representation
character which is obtained fron the expression  given in the Appendix
after the replacement $z_i \to z^2_i$.

Using similar prescription for the magnetic theory, we find
\beqa
&& I_M \ = \ \kappa_4 \int_{\mathbb{T}^4} \prod_{1 \leq i < j \leq 4}
\frac{\prod_{k=1}^3 \Gamma(s_k
y_i^{\pm1}y_j^{\pm1};p,q)}{\Gamma(y_i^{\pm1}y_j^{\pm1};p,q)}
\prod_{j=1}^4 \frac{\prod_{k=1}^3 \Gamma(s_k
y_j^{\pm2};p,q)}{\Gamma(y_j^{\pm2};p,q)} \nonumber
\\ && \makebox[5em]{} \times
\frac{\prod_{k=1}^3 \Gamma(s_k
y_1^{\pm1}y_2^{\pm1}y_3^{\pm1}y_4^{\pm1};p,q)}{\Gamma(y_1^{\pm1}y_2^{\pm1}y_3^{\pm1}y_4^{\pm1};p,q)}
\prod_{j=1}^4 \frac{dy_j}{2 \pi \textup{i} y_j}.\eeqa
These are the first examples
of multiple elliptic hypergeometric integrals defined for the $F_4$
root system (in \cite{Bult} the integrals were defined on the
$SU(2)$ group and the Weyl group $W(F_4)$ was acting in the parameter space).

The $S$-duality conjecture suggests the transformation formula $I_E = I_M$.
Again, as suggested to us by S. Razamat, this identity is
easily established by the change of variables
$$
y_1  =z_1z_2,\qquad  y_2 = z_1/z_2, \qquad y_3=z_3z_4,\qquad
y_4=z_3/z_4,
$$
associated with the rotation of the $F_4$ root system \cite{AKS}.
We see thus validity of the SCI test for this $S$-duality.

The limit $p=s_2=0$ reduces integral (\ref{F4}) to
\beqa  \nonumber
&& I_{F_4}^{p=s_2=0}  = \frac{1}{2^{7} 3^2}
\frac{(q;q)^4_\infty}{(s_1;q)_\infty^4}
\int_{\mathbb{T}^4} \prod_{1 \leq i < j \leq 4}
\frac{(z_i^{\pm2}z_j^{\pm2};q)_\infty}{(s_1z_i^{\pm2}z_j^{\pm2};q)_\infty}
\prod_{j=1}^4 \frac{(z_j^{\pm2};q)_\infty}{(s_1z_j^{\pm2};q)_\infty}
\\ && \makebox[5em]{} \times
\frac{(z_1^{\pm1}z_2^{\pm1}z_3^{\pm1}z_4^{\pm1};q)_\infty}
{(s_1z_1^{\pm1}z_2^{\pm1}z_3^{\pm1}z_4^{\pm1};q)_\infty}
\prod_{j=1}^4 \frac{dz_j}{2 \pi \textup{i} z_j},
\label{F4_lim3d}\eeqa
which admits exact evaluation \cite{Ito}
\beqa
I_{F_4}^{p=s_2=0} \ = \
\frac{(qs_1,qs_1^5,qs_1^7,qs_1^{11};q)_\infty}
{(s_1^2,s_1^6,s_1^8,s_1^{12};q)_\infty}.\eeqa

\section{$SU(N)$ and $SO(2N)$ gauge groups}

Consider now SCIs for self-dual $\mathcal{N}=4$ SYM theories
with $SU(N)$ and $SO(2N)$ gauge groups \cite{Goddard}.
The $SU(N)$ theory SCI is
 \beq \label{SU}
I_{SU(N)} = \chi_N \int_{\mathbb{T}^{N-1}} \prod_{1 \leq i < j \leq
N} \frac{\prod_{k=1}^3 \Gamma(s_k z_i^{-1}z_j, s_k
z_iz_j^{-1};p,q)}{\Gamma(z_i^{-1}z_j,z_iz_j^{-1};p,q)}
\prod_{j=1}^{N-1} \frac{dz_j}{2 \pi \textup{i} z_j},\eeq
where $\prod_{j=1}^N z_j = 1$, parameters $s_k$ satisfy the constraints
$|s_k|<1,\, k=1,2,3,$ and
$$\chi_N \ = \ \frac{(p;p)^{N-1}_\infty (q;q)^{N-1}_\infty}{N!}
\prod_{k=1}^3 \Gamma^{N-1}(s_k;p,q).$$

The limit $p=0, s_2=0$ reduces integral  (\ref{SU}) to
\beq \label{SU_lim3d}
I_{SU(N)}^{p=s_2=0} = \frac{1}{N!}
\frac{(q;q)^{N-1}_\infty}{(s_1;q)_\infty^{N-1}}
\int_{\mathbb{T}^{N-1}} \prod_{1 \leq i < j \leq N}
\frac{(z_i^{-1}z_j,z_iz_j^{-1};q)_\infty}{(s_1 z_i^{-1}z_j, s_1
z_iz_j^{-1};q)_\infty} \prod_{j=1}^{N-1} \frac{dz_j}{2 \pi
\textup{i} z_j}, \eeq where $\prod_{j=1}^N z_j=1$,
which admits exact evaluation \cite{Ito}
\beqa \label{SU_ans}
I_{SU(N)}^{p=s_2=0} \ = \ \prod_{j=1}^{N-1}
\frac{(qs_1^j;q)_\infty}{(s_1^{j+1};q)_\infty}.\eeqa
For $N\to\infty$ this index equals to $(s_1;q)_\infty/(s_1;s_1)_\infty$,
which coincides with the reduced form of $N\to\infty$ asymptotics
(after passing from $U(N)$ to $SU(N)$ gauge group)
found in \cite{Kinney} from the AdS/CFT correspondence.

SCI for the $SO(2N)$ theory has the form
\beq \label{SO}
I_{SO(2N)} = \chi_N \int_{\mathbb{T}^{N}} \prod_{1 \leq i < j \leq
N} \frac{\prod_{k=1}^3 \Gamma(s_k
z_i^{\pm1}z_j^{\pm1};p,q)}{\Gamma(z_i^{\pm1}z_j^{\pm1};p,q)}
\prod_{j=1}^{N}\frac{dz_j}{2 \pi \textup{i} z_j}, \eeq
where $|s_k|<1,\, k=1,2,3,$ and
$$
\chi_N \ = \ \frac{(p;p)^{N}_\infty (q;q)^{N}_\infty}{2^{N-1} N!}
\prod_{k=1}^3 \Gamma^{N}(s_k;p,q).$$
Note that for $N=1$ the SCI is equal to $\chi_1$.

Taking the ratio of integral kernel to itself with
different integration variables in \eqref{SU} and \eqref{SO}
 one gets totally elliptic
hypergeometric terms. However, consequences of this statement are
much less informative than in the cases with nontrivial
symmetry transformations for integrals.

The limit $p=0, s_2=0$ reduces (\ref{SO}) to the integral
\beq \label{SO2n_lim3d}
I_{SO(2N)}^{p=s_2=0} = \frac{1}{2^{N-1}N!}
\frac{(q;q)^{N}_\infty}{(s_1;q)_\infty^{N}}
 \int_{\mathbb{T}^{N}} \prod_{1 \leq i < j \leq N}
\frac{(z_i^{\pm1}z_j^{\pm1};q)_\infty}{(s_1z_i^{\pm1}z_j^{\pm1};q)_\infty}
\prod_{j=1}^{N} \frac{dz_j}{2 \pi \textup{i} z_j}, \eeq
 with exact evaluation \cite{Ito}
\beqa
I_{SO(2N)}^{p=s_2=0} \ = \
\frac{(qs_1^{N-1};q)_\infty}{(s_1^N;q)_\infty} \prod_{j=0}^{N-2}
\frac{(qs_1^{2j+1};q)_\infty}{(s_1^{2j+2};q)_\infty}.\eeqa
In the same way as for $SP(2N)$ and $SO(2N+1)$ SYM theories, this case can be
obtained from the $q$-Selberg integral (\ref{gus}) using
special parameter values
$$
b=s_1,\qquad a_{1,2}=\pm 1,\qquad a_{3,4}=\pm \sqrt{q}.
$$

Consider now the hyperbolic degeneration of \eqref{SU} and \eqref{SO} joint with
the $\alpha_2 \to \infty$ limit similar to $SP(2N)$ and $SO(2N+1)$ SCIs.
 For $SU(N)$-SCI we obtain, after dropping the multiplier
$\exp \{\frac{\pi \textup{i}}{2}(\alpha_1^2 + 2 \alpha_1 \alpha_2 - \alpha_1
(\omega_1+\omega_2)) (N^2-1)\}$,
\beq \label{hyperSUNe}
I_{SU(N)}^{h,\alpha_2 \to \infty} =
\frac{\gamma^{(2)}(\alpha_1;\omega)^{N-1}}{N!(\textup{i}\sqrt{\omega_1\omega_2})^{N-1}}
 \int_{- \textup{i} \infty}^{\textup{i} \infty}
\prod_{1 \leq i < j \leq N} \frac{\gamma^{(2)}( \alpha_1 \pm (u_i - u_j);\mathbf{\omega})}
{\gamma^{(2)}( \pm (u_i - u_j);\mathbf{\omega})} \prod_{j=1}^{N-1} du_j,
\eeq
where $\sum_{j=1}^N u_j = 0$. In the analysis of convergency of this integral
there are two extremal options when
integration variables go to infinity: in the first case
$u_j = \textup{i} R + v_j, j=1,\ldots,N-1$, and
$u_N = - (N-1) \textup{i} R - \sum_{j=1}^{N-1} v_j$, where
$R \rightarrow +\infty$, and the integrand behaves as
$\exp (2 \pi N (N-1) \alpha_1 R/\omega_1\omega_2)$.
In the second case, $u_1 = \textup{i} R, \Im(u_j) \ll R, j=2,\ldots,N-1$,
and $u_N = - \textup{i} R - \sum_{j=2}^{N-1} u_j$,
 $R \rightarrow +\infty$, and the integrand behaves as
$\exp (2 \pi N \alpha_1 R/\omega_1\omega_2)$.
In both cases, for $\Re(\alpha_1/\omega_1\omega_2) < 0$
the integrand is exponentially suppressed and
has no singularities on the integration contour.

To our knowledge integral \eqref{hyperSUNe} cannot be obtained
as a limit of known hyperbolic beta integrals. Formally it is
related to the limit $\sum_{i=1}^4 \mu_i+(N-1) \tau -\omega_1-\omega_2
\to 0$ in formula \eqref{conf2}, which is not uniform.
Therefore we have separately computed this integral for $N=2, 3$
by showing that the sum of residues for poles
on the left-hand side of the integration contours is proportional
to the product of sums of residues of two trigonometric integrals
\eqref{SU_ans} with bases $q=e^{2\pi\textup{i}\omega_1/\omega_2}$
and $\tilde q=e^{-2\pi\textup{i} \omega_2/\omega_1}$, $|q|<1$,
which yields
\beq \label{hyperSUN}
I_{SU(N)}^{h,\alpha_2 \rightarrow \infty}
= \prod_{j=1}^{N-1} \frac{\gamma^{(2)}((j+1) \alpha_1;\mathbf{\omega})}
{\gamma^{(2)}(j\alpha_1 + \omega_1 + \omega_2;\mathbf{\omega})}.
\eeq
For $N=4$ this integral coincides with the $SO(6)$-integral given below.
Note that formula \eqref{hyperSUN} defines a hyperbolic analogue of
the orthogonality measure normalization for Macdonald polynomials
on $A_{N-1}$ root system \eqref{SU_lim3d}, \eqref{SU_ans}
(for arbitrary $N$ we consider it as a conjecture).

The hyperbolic limit for SCI of $SO(2N)$-theory ($N>1$) yields,
after dropping the multiplier
$\exp \{\frac{\pi \textup{i}}{2} (\alpha_1^2 + 2 \alpha_1 \alpha_2 - \alpha_1
(\omega_1+\omega_2)) (2N^2-N)\}$,
\beq \label{hyperSO2Ne}
I_{SO(2N)}^{h,\alpha_2 \to \infty} = \xi_N \int_{- \textup{i} \infty}^{\textup{i} \infty}
\prod_{1 \leq i < j \leq N} \frac{\gamma^{(2)}( \alpha_1 \pm u_i \pm u_j;\mathbf{\omega})}
{\gamma^{(2)}( \pm u_i \pm u_j;\mathbf{\omega})} \prod_{j=1}^N du_j.
\eeq
This integral is obtained from (\ref{conf2}) after the substitutions
$$
\tau = \alpha_1, \ \mu_1 = 0, \ \mu_2 = \frac 12 \omega_1, \ \mu_3 = \frac 12 \omega_2, \
\mu_4 = \frac 12 (\omega_1+\omega_2),$$
which leads to the evaluation
\beq \label{hyperSO2N}
I_{SO(2N)}^{h,\alpha_2 \rightarrow \infty}
= \frac{\gamma^{(2)}(N \alpha_1;\mathbf{\omega})}
{\gamma^{(2)}((N-1) \alpha_1 + \omega_1 + \omega_2;\mathbf{\omega})}
 \prod_{j=0}^{N-2} \frac{\gamma^{(2)}(2 (j+1) \alpha_1;\mathbf{\omega})}
{\gamma^{(2)}((2j+1) \alpha_1 + \omega_1 + \omega_2;\mathbf{\omega})}.
\eeq

Again, one can see that expressions \eqref{hyperSUNe} and \eqref{hyperSUN},
\eqref{hyperSO2Ne} and \eqref{hyperSO2N} describe
partition functions of $3d$ $\mathcal{N}=2$ SYM theories with one chiral matter
superfield in the adjoint representation of the respective $SU(N)$
and $SO(2N)$ gauge groups and their dual confining partners.
Substitution $\alpha_1 = (\omega_1+\omega_2)/2$ in these expressions
leads to vanishing partition functions of $3d$ $\mathcal{N}=4$ pure SYM theories.

\section{Exceptional gauge groups $E_6, E_7,$ and $E_8$}

For the $E_6$ gauge group theory we have the SCI
\beqa \label{E6} && \makebox[-2em]{}
I_{E_6}  = \kappa_6
\int_{\mathbb{T}^6} \prod_{j=1}^6 \frac{dz_j}{2 \pi \textup{i} z_j}
\prod_{1 \leq i < j \leq 5} \frac{\prod_{k=1}^3 \Gamma(s_k
z_i^{\pm2}z_j^{\pm2};p,q)}{\Gamma(z_i^{\pm2}z_j^{\pm2};p,q)}
\frac{\prod_{k=1}^3 \Gamma(s_k
(z_6^{3}Z)^{\pm1};p,q)}{\Gamma((z_6^{3}Z)^{\pm1};p,q)} \nonumber
\\ && \makebox[-1em]{} \times \prod_{1 \leq i < j \leq 5}
\frac{\prod_{k=1}^3 \Gamma(s_k (z_6^{3} z_i^2z_j^2
Z)^{\pm1};p,q)}{\Gamma((z_6^{3} z_i^2z_j^2Z)^{\pm1};p,q)}
\prod_{i=1}^5 \frac{\prod_{k=1}^3 \Gamma(s_k (z_6^{-3}
z_i^{2}Z)^{\pm1};p,q)}{\Gamma((z_6^{-3} z_i^{2}Z)^{\pm1};p,q)},
\eeqa
where for convenience we denoted
$Z = (z_1z_2z_3z_4z_5)^{-1}$  and
$$
\kappa_6 \ = \ \frac{(p;p)^6_\infty (q;q)^6_\infty}{2^7 3^4 5}
\prod_{k=1}^3 \Gamma^{6}(s_k;p,q).
$$
The combinatorial factors appearing here are the same as, for example,
the ones given in \cite{Ito}. Similar to the $F_4$-group case, we took
the adjoint representation character given in the Appendix and replaced
in it $z_j\to z_j^2$ (the same was done for the $E_7$ and $E_8$ group
cases considered below).

The limit $p=0, s_2=0$ reduces (\ref{E6}) to the integral
\beqa \label{E6_lim3d} && I_{E_6}^{p=s_2=0}  =\frac{1}{2^{7} 3^4 5}
\frac{(q;q)^6_\infty}{(s_1;q)_\infty^6}
\int_{\mathbb{T}^6} \prod_{j=1}^6 \frac{dz_j}{2 \pi
\textup{i} z_j} \prod_{1 \leq i < j \leq 5}
\frac{(z_i^{\pm2}z_j^{\pm2};q)_\infty}{(s_1z_i^{\pm2}z_j^{\pm2};q)_\infty}
\nonumber
\\ && \makebox[-2em]{} \times
\frac{((z_6^{3}Z)^{\pm1};q)_\infty}{(s_1(z_6^{3}Z)^{\pm1};q)_\infty}
\prod_{1 \leq i < j \leq 5}
\frac{((z_6^{3} z_i^2z_j^2 Z)^{\pm1};q)_\infty}{(s_1(z_6^{3}
z_i^2z_j^2Z)^{\pm1};q)_\infty} \prod_{i=1}^5 \frac{((z_6^{-3}
z_i^{2}Z)^{\pm1};q)_\infty}{(s_1(z_6^{-3}
z_i^{2}Z)^{\pm1};q)_\infty},\eeqa
which can be computed exactly \cite{Ito},
\beqa
I_{E_6}^{p=s_2=0} \ = \
\frac{(qs_1,qs_1^4,qs_1^5,qs_1^7,qs_1^8,qs_1^{11};q)_\infty}
{(s_1^2,s_1^5,s_1^6,s_1^8,s_1^9,s_1^{12};q)_\infty}.\eeqa

For $\mathcal{N}=4$ SYM theory with the $E_7$ gauge group the SCI has the form
\beqa \label{E7} && \makebox[-2em]{}
I_{E_7} = \kappa_7 \int_{\mathbb{T}^7}
\prod_{j=1}^6 \frac{\prod_{k=1}^3 \Gamma(s_k z_7^{\pm2}
(z_j^2Z)^{\pm1};p,q)}{\Gamma(z_7^{\pm2} (z_j^2Z)^{\pm1};p,q)}
\prod_{1 \leq i < j \leq 6} \frac{\prod_{k=1}^3 \Gamma(s_k
z_i^{\pm2}z_j^{\pm2};p,q)}{\Gamma(z_i^{\pm2}z_j^{\pm2};p,q)} \nonumber \\
&& \makebox[-1em]{} \times \frac{\prod_{k=1}^3 \Gamma(s_k
z_7^{\pm4};p,q)}{\Gamma(z_7^{\pm4};p,q)}\prod_{1 \leq i < j < l \leq
6} \frac{\prod_{k=1}^3 \Gamma(s_k z_7^{\pm2}
z_i^2z_j^2z_l^2Z;p,q)}{\Gamma(z_7^{\pm2} z_i^2z_j^2z_l^2Z;p,q)}
\prod_{j=1}^7 \frac{dz_j}{2 \pi \textup{i} z_j},
\eeqa
where we denoted $Z = (z_1z_2z_3z_4z_5z_6)^{-1}$ and
$$
\kappa_7 \ = \ \frac{(p;p)^7_\infty (q;q)^7_\infty}{2^{10} 3^4 5 \cdot 7}
\prod_{k=1}^3 \Gamma^{7}(s_k;p,q).
$$

The limit $p=0, s_2=0$ reduces (\ref{E7}) to the
integral \beqa \label{E7_lim3d} &&\makebox[-2em]{}
 I_{E_7}^{p=s_2=0} =
\frac{1}{2^{10} 3^4 5 \cdot 7}
\frac{(q;q)^7_\infty}{(s_1;q)_\infty^7}
 \int_{\mathbb{T}^7} \prod_{j=1}^6 \frac{(z_7^{\pm2}
(z_j^2Z)^{\pm1};q)_\infty}{(s_1z_7^{\pm2} (z_j^2Z)^{\pm1};q)_\infty}
\prod_{1 \leq i < j \leq 6} \frac{(z_i^{\pm2}z_j^{\pm2};q)_\infty}{(s_1z_i^{\pm2}z_j^{\pm2};q)_\infty} \nonumber \\
&& \makebox[2em]{} \times
\frac{(z_7^{\pm4};q)_\infty}{(s_1z_7^{\pm4};q)_\infty} \prod_{1 \leq
i < j < l \leq 6} \frac{(z_7^{\pm2} z_i^2z_j^2z_l^2Z;q)_\infty}{(s_1
z_7^{\pm2} z_i^2z_j^2z_l^2Z;q)_\infty} \prod_{j=1}^7 \frac{dz_j}{2
\pi \textup{i} z_j},\eeqa
which can be evaluated exactly \cite{Ito},
\beqa I_{E_7}^{p=s_2=0} \ = \
\frac{(qs_1,qs_1^5,qs_1^7,qs_1^9,qs_1^{11},qs_1^{13},qs_1^{17};q)_\infty}
{(s_1^2,s_1^6,s_1^8,s_1^{10},s_1^{12},s_1^{14},s_1^{18};q)_\infty}.\eeqa

Finally,  the largest exceptional gauge group $E_8$ theory has the SCI
\beqa \label{E8} &&
\makebox[-2em]{} I_{E_8}  = \kappa_8 \int_{\mathbb{T}^8}
\prod_{j=1}^8 \frac{dz_j}{2 \pi \textup{i} z_j} \prod_{1 \leq i < j
\leq 8} \frac{\prod_{k=1}^3 \Gamma(s_k (z_i^2z_j^2Z)^{\pm
1};p,q)}{\Gamma((z_i^2z_j^2Z)^{\pm 1};p,q)} \frac{\prod_{k=1}^3
\Gamma(s_k Z^{\pm1};p,q)}{\Gamma(Z^{\pm1};p,q)} \\ \nonumber &&
\makebox[-2em]{} \times  \prod_{1 \leq i < j \leq 8}
\frac{\prod_{k=1}^3 \Gamma(s_k
z_i^{\pm2}z_j^{\pm2};p,q)}{\Gamma(z_i^{\pm2}z_j^{\pm2};p,q)}\prod_{1
\leq i < j < l < m \leq 8} \frac{\prod_{k=1}^3 \Gamma(s_k
(z_i^2z_j^2z_l^2z_m^2Z)^{\pm
1};p,q)}{\Gamma((z_i^2z_j^2z_l^2z_m^2Z)^{\pm 1};p,q)},\eeqa where
$Z= (z_1z_2z_3z_4z_5z_6z_7z_8)^{-1}$ and
$$\kappa_8 \ = \ \frac{(p;p)^8_\infty (q;q)^8_\infty}{2^{14} 3^5 5^2 7}
\prod_{k=1}^3 \Gamma^{8}(s_k;p,q).$$

Again, the limit $p=0, s_2=0$ reduces (\ref{E8}) to the integral
 \beqa \label{E8_lim3d} && \makebox[-2em]{}
I_{E_8}^{p=s_2=0} =
\frac{1}{2^{14} 3^5 5^2 7}
\frac{(q;q)^8_\infty}{(s_1;q)_\infty^8} \int_{\mathbb{T}^8} \prod_{j=1}^8 \frac{dz_j}{2 \pi
\textup{i} z_j} \prod_{1 \leq i < j \leq 8}
\frac{((z_i^2z_j^2Z)^{\pm 1};q)_\infty}{(s_1(z_i^2z_j^2Z)^{\pm
1};q)_\infty}
\\ \nonumber && \makebox[-1em]{} \times
 \frac{(Z^{\pm1};q)_\infty}{(s_1 Z^{\pm1};q)_\infty}
 \prod_{1 \leq i < j \leq 8}
\frac{(z_i^{\pm2}z_j^{\pm2};q)_\infty}{(s_1
z_i^{\pm2}z_j^{\pm2};q)_\infty} \prod_{1 \leq i < j < l < m \leq 8}
\frac{((z_i^2z_j^2z_l^2z_m^2Z)^{\pm 1};q)_\infty}{(s_1
(z_i^2z_j^2z_l^2z_m^2Z)^{\pm 1};q)_\infty},\eeqa
which can be evaluated exactly \cite{Ito},
 \beqa
I_{E_8}^{p=s_2=0} \ = \
\frac{(qs_1,qs_1^7,qs_1^{11},qs_1^{13},qs_1^{17},qs_1^{19},qs_1^{23},qs_1^{29};q)_\infty}{
(s_1^2,s_1^8,s_1^{12},s_1^{14},s_1^{18},s_1^{20},s_1^{24},s_1^{30};q)_\infty}.
\eeqa

In all three integrals \eqref{E6}, \eqref{E7}, and \eqref{E8}
 we assumed the restrictions $|s_k|<1,\, k=1,2,3.$
As expected, ratios of their kernels to themselves with different
integration variables yield totally elliptic hypergeometric terms.
These integrals represent first known examples of elliptic
hypergeometric integrals based on the exceptional root systems of
$E$--type.

\section{Some special ${\mathcal N}=1$ and ${\mathcal N}=2$ dualities}

Much attention is paid in this paper to supersymmetric theories with
the exceptional gauge groups. Therefore we would like to describe one
more duality example for such theories known to us.
We take  ${\mathcal N}=1$ SYM theory
with $E_6$ gauge group and matter fields in the
fundamental representation of $SU(6)$ flavor group and in
$27$-dimensional representation of $E_6$.

This electric theory and its magnetic dual were suggested in
\cite{Exc1,Exc2} and validity of this duality was discussed further in
 \cite{Pexc}. The electric SCI is
\beqa &&\makebox[-2em]{}
I_{E} = \kappa_6 \int_{\mathbb{T}^6}\prod_{1 \leq i < j
\leq 5} \frac{\prod_{k=1}^6 \Gamma(s_kz_6^{-1} Z z_i^{2}
z_j^2;p,q)}{\Gamma(z_i^{\pm2}z_j^{\pm2};p,q)} \frac{\prod_{k=1}^6
\Gamma(s_kz_6^{-4}, s_k
z_6^{-1}Z;p,q)}{\Gamma((z_6^{3}Z)^{\pm1};p,q)}
\\  \nonumber && \makebox[-2em]{} \times \prod_{1 \leq i < j \leq 5}
\frac{1}{\Gamma((z_6^{3} z_i^2z_j^2Z)^{\pm1};p,q)} \prod_{i=1}^5
\frac{\prod_{k=1}^6 \Gamma(s_kz_6^2 z_i^{\pm2},
s_kz_6^{-1}Z^{-1}z_i^{-2};p,q)}{\Gamma((z_6^{-3}
z_i^{2}Z)^{\pm1};p,q)}\prod_{j=1}^6 \frac{dz_j}{2 \pi \textup{i} z_j} ,\eeqa
where $|s_k|<1,\, k=1,\ldots,6,$
we denoted $Z = (z_1z_2z_3z_4z_5)^{-1}$ and
$$\kappa_6 \ = \ \frac{(p;p)^6_\infty (q;q)^6_\infty}{2^7 3^4 5}.$$

 The magnetic theory has chiral fields in the antifundamental representation
of the flavor group and $27$-dimensional representation of the gauge
group. There are also singlet mesons given by the
absolutely symmetric representation of the third rank of the flavor group.
The magnetic SCI is
\beqa
&&\makebox[-2em]{} I_{M} = \kappa_6 \prod_{j=1}^6 \Gamma(s_j^3;p,q)
\prod_{i,j=1;\, i \neq j}^6 \Gamma(s_is_j^2;p,q) \int_{\mathbb{T}^6}
\prod_{1 \leq i < j  \leq 5}
\frac{1}{\Gamma((z_6^{3} z_i^2z_j^2Z)^{\pm1};p,q)}
\nonumber \\
&& \makebox[-1em]{} \times \prod_{1 \leq i < j \leq 5}
\frac{\prod_{k=1}^6 \Gamma(S^{\frac 13}s^{-1}_k z_6^{-1} Z
z_i^{2}z_j^2;p,q)}{\Gamma(z_i^{\pm2}z_j^{\pm2};p,q)}
\frac{\prod_{k=1}^6 \Gamma(S^{\frac 13}s^{-1}_kz_6^{-4}, S^{\frac
13}s^{-1}_k z_6^{-1}Z;p,q)}{\Gamma((z_6^{3}Z)^{\pm1};p,q)} \nonumber
\\ && \makebox[-1em]{} \times \prod_{i=1}^5
\frac{\prod_{k=1}^6 \Gamma(S^{\frac 13}s^{-1}_kz_6^2 z_i^{\pm2},
S^{\frac 13}s^{-1}_k z_6^{-1}Z^{-1}z_i^{-2};p,q)}{\Gamma((z_6^{-3}
z_i^{2}Z)^{\pm1};p,q)}\prod_{j=1}^6 \frac{dz_j}{2 \pi \textup{i} z_j} ,\eeqa
where $|s_k|<1, \, k=1,\ldots,6$.
The balancing condition for both elliptic hypergeometric
integrals  has the form  $S=\prod_{i=1}^6 s_i = pq$.

We have checked that the ratio of these integral kernels yields a
totally elliptic hypergeometric term, which is an important test suggesting
that these dualities and the equality $I_E=I_M$ might be true.
Interestingly, the limit $s_6 \to 1$ reduces the integrals to SCIs of
peculiar $E_6$ and $F_4$ SYM theories dual to each other \cite{Exc2}.

Finally, as an additional advertisement of the applications of the theory
of elliptic hypergeometric integrals, we present SCI of a
particular $\mathcal{N}=2$ quiver SYM theory described in
\cite{Tachikawa:2009rb}. Define
\beqa && I_E \ = \
\frac{(p;p)^6_\infty(q;q)^6_\infty}{8} \int_{\mathbb{T}} \frac{dx}{2
\pi \textup{i} x} \int_{\mathbb{T}} \frac{dy}{2 \pi \textup{i} y}
\int_{\mathbb{T}^2} \prod_{j=1}^2 \frac{dz_j}{2 \pi \textup{i} z_j}
\int_{\mathbb{T}} \frac{dr}{2 \pi \textup{i} r} \int_{\mathbb{T}}
\frac{dw}{2 \pi \textup{i} w} \nonumber \\ && \makebox[2em]{} \times
\frac{\Gamma(t^2vx^{\pm1},t^2vy^{\pm2},t^2vz_1^{\pm1}z_2^{\pm1},t^2vr^{\pm2},t^2vw^{\pm1};p,q)}
{\Gamma(x^{\pm1},y^{\pm2},z_1^{\pm1}z_2^{\pm1},r^{\pm2},w^{\pm1};p,q)}
\nonumber \\
&& \makebox[2em]{} \times
\Gamma\Big(\frac{t^2}{\sqrt{v}} y^{\pm1},\frac{t^2}{\sqrt{v}}
r^{\pm1};p,q\Big)^2
\Gamma\Big(\frac{t^2}{\sqrt{v}} x^{\pm1}y^{\pm1},\frac{t^2}{\sqrt{v}}
r^{\pm1}w^{\pm1};p,q\Big) \nonumber \\
&& \makebox[2em]{} \times \prod_{j=1}^2 \Gamma\Big(\frac{t^2}{\sqrt{v}}
y^{\pm1}z_j^{\pm1},\frac{t^2}{\sqrt{v}} r^{\pm1}z_j^{\pm1};p,q\Big),
\eeqa
where $t$ is the same parameter as in ${\mathcal N}=4$ theories
before and $v$ is the chemical potential associated with some
combination of the $U(2)_R$-group commuting $R$-charges. Introducing the variables
$\alpha^2 =  z_1z_2, \ \beta^2  = z_1/z_2,\  \gamma^2 =x$, and $\delta^2  =  w$,
one can rewrite this integral as
\beqa \nonumber &&
\makebox[-1em]{} I_M \ = \ \frac{(p;p)^6_\infty(q;q)^6_\infty}{64}
\int_{\mathbb{T}}
\frac{d\gamma}{2 \pi \textup{i} \gamma} \int_{\mathbb{T}} \frac{dy}{2 \pi \textup{i}
y} \int_{\mathbb{T}} \frac{d\alpha}{2 \pi \textup{i} \alpha}
\int_{\mathbb{T}} \frac{d\beta}{2 \pi \textup{i} \beta} \int_{\mathbb{T}}
\frac{dr}{2 \pi \textup{i} r} \int_{\mathbb{T}} \frac{d\delta}{2 \pi \textup{i}
\delta}
\\ && \makebox[0em]{} \times
\frac{\Gamma(t^2v\gamma^{\pm2},t^2vy^{\pm2},t^2v\alpha^{\pm2},t^2v\beta^{\pm2},
t^2vr^{\pm2},t^2v\delta^{\pm2};p,q)}
{\Gamma(\gamma^{\pm2},y^{\pm2},\alpha^{\pm2},\beta^{\pm2},r^{\pm2},\delta^{\pm2};p,q)}
\\ \nonumber && \makebox[-2em]{} \times
\Gamma\Big(\frac{t^2}{\sqrt{v}}
\gamma^{\pm1}\gamma^{\pm1}y^{\pm1},\frac{t^2}{\sqrt{v}}
\delta^{\pm1}\delta^{\pm1}r^{\pm1},\frac{t^2}{\sqrt{v}}
y^{\pm1}\alpha^{\pm1}\beta^{\pm1},\frac{t^2}{\sqrt{v}}
r^{\pm1}\alpha^{\pm1}\beta^{\pm1};p,q\Big). \eeqa

The identity $I_E = I_M$ can be interpreted as the equality of
SCIs for particular $\mathcal{N}=2$ SYM generalized quiver theories
(although it does not correspond to an intrinsic electric-magnetic duality).
The ``electric" part is an
$SO(3) \times SP(2) \times SO(4) \times SP(2) \times SO(3)$ $\mathcal{N}=2$
SYM quiver and the ``magnetic" part is the same theory rewritten as an
$SU(2)^6$-quiver, as illustrated in Fig. 9 of \cite{Tachikawa:2009rb}.

\section{Discussion}

In this paper we have described SCIs for $\mathcal{N}=4$ SYM theories with
simple gauge groups as elliptic hypergeometric integrals
and analyzed some of their mathematical properties.
For all classical simple gauge groups we have found particular limiting
values of chemical potentials ($p\to 0$ followed by the $s_2\to 0$ limit
and the hyperbolic limit followed by the $\alpha_2\to\infty$ limit)
for which ${\mathcal N}=4$ indices are computable
exactly. According to the general ideology \cite{Romelsberger1,Dolan:2008qi,SV2},
exact computability of non-abelian gauge group SCIs is associated with the confinement
in the dual phase of the theory, since it provides a group-theoretical
representation of indices without local gauge group symmetry.
Therefore we conclude that there should exist some
interesting supersymmetric field theories
similar to the Wess-Zumino model whose SCIs are described by the
right-hand sides of equalities (\ref{SP_lim3d}), (\ref{SO_lim3d}),
(\ref{G2_1_lim3d}), (\ref{F4_lim3d}), (\ref{SU_lim3d}),
(\ref{SO2n_lim3d}), (\ref{E6_lim3d}), (\ref{E7_lim3d}), and
(\ref{E8_lim3d}). The hyperbolic analogs of these relations describe equalities
of $3d$ partition functions of particular dual $3d$  $\mathcal{N}=2$
and $\mathcal{N}=4$ SYM theories.

To our knowledge, hyperbolic beta integrals for exceptional
groups were not considered in the literature. Analysing
such exact integration formulas given in \cite{S3,ds,rai:limits,DSV} and references
therein, we conjecture that the hyperbolic analogs of all our
exceptional gauge group $q$-beta integrals are obtained from
them after the replacement of infinite products
$(q^ns_k^mz_j^\ell;q)_\infty$ with $m$ or $\ell\neq 0$ by $1/\gamma^{(2)}(n(\omega_1+\omega_2)
+m\alpha_k+\ell u_j;\mathbf{\omega}),$
the measure elements $(q;q)_\infty dz_j/2\pi\textup{i}z_j$
by $du_j/\textup{i}\sqrt{\omega_1\omega_2},$
and $\mathbb T$ by the Mellin-Barnes integration contours.
From the physical point of view this is equivalent to the conjecture on
the particular structure of confining phases of corresponding $3d$ $\mathcal{N}=2$ SYM
theories with $G_2, F_4, E_6, E_7, E_8$ gauge groups and one matter field in the
adjoint representation. For $\alpha_1 = (\omega_1+\omega_2)/2$ this would yield
vanishing partition functions for $3d$ $\mathcal{N}=4$ pure SYM theories.

One of the initial motivations for consideration of SCIs
in \cite{Kinney} was an analysis of the AdS/CFT
correspondence for $\mathcal{N}=4$ SYM theory with
$U(N)$ gauge group which required  consideration of the $N\to\infty$ limit.
In this limit, the original index coming from the BPS states  not forming long multiplets
can be computed from the dual spectrum of gravitons
appearing in the Type IIB supergravity compactified on $AdS_5 \times  S^5$.
It would be interesting to understand the meaning of the reduction
$p \rightarrow 0$ from the AdS/CFT point of view on the level
of graviton spectra. All our $p=s_2=0$ indices for gauge groups of rank
$N$ are well defined in the limit $N\to\infty$ for
$|s_1|<1$, being given by curious explicit infinite products.
We expect that the $p=s_2=0$ limit corresponds to an essentially simplified
picture for the corresponding gravitational duals for both finite
and infinite $N$.

In \cite{Kol,Seiberg2010}, marginal deformations of SCFTs
were studied and the importance of global symmetries
for the conformal manifold (a manifold of coupling constants of the theory
where it stays conformal) is shown.
A $\beta$-deformation of the ${\mathcal N}=4$ SYM
theory \cite{Lunin:2005jy} is obtained by introduction of a marginal
deformation of the superpotential $h\text{Tr}( e^{\textup{i} \pi \beta} \Phi_1
\Phi_2 \Phi_3 - e^{-\textup{i} \pi \beta} \Phi_1 \Phi_3 \Phi_2)$
breaking  $\mathcal{N}=4$ supersymmetry down to $\mathcal{N}=1$
($h$ is the Yukawa coupling). The arbitrary parameter $\beta$ may be complex
and this does not spoil superconformal invariance of the theory
\cite{Kazakov:2007dy}. The initial $R$-symmetry $SU(4)_R$
breaks to $U(1)_R$ with the additional global symmetry $U(1)_1 \times
U(1)_2$ \cite{Lunin:2005jy}. From the indices point of view the
parameters $v$ and $w$ play now the role of chemical potentials
for the latter global group.
SCI for the $\beta$-deformed theory is the same as in the initial
theory \cite{Kinney}. This means that these theories share
essentially the same set of BPS states.
In the conclusion of \cite{SV2}, we discussed appearance of
the $SO(3)$ $\mathcal{N}=4$ SYM theory from an $\mathcal{N}=1$ model after
a superpotential deformation, such that both theories share the same SCI.
Actually, SCIs of all exactly marginally deformed theories coincide,
only the interpretation of chemical potentials is different, being tied to
global groups of different meaning. Therefore these indices serve as invariants
of the conformal manifold with their structure reflecting only
a part of the global symmetries preserved by the superpotential.

As an example of different deformation of $\mathcal{N}=4$ theories we can
mention the deformation to  $\mathcal{N}=1$ SYM theory with two chiral superfields
in the adjoint representation and an additional $U(1)$ global group
(see \cite{Argyres} and references therein).
This theory has an $SL(2,\mathbb{Z})$ group electric-magnetic
duality inherited from $\mathcal{N}=4$ SYM theory in its infrared fixed point.
At the level of SCIs such a deformation is realized in a very simple way,
it is just necessary to fix, say, $s_3= \sqrt{pq}$, which excludes
this parameter completely from the integrals.

The $q$-beta integrals appearing from SCIs of all ${\mathcal N}=4$ SYM
theories in the limit $p\to0$, $s_2\to 0$
determine orthogonality measures for special cases of the Koornwinder
and Macdonald orthogonal polynomials
(for $E_6, E_7,$ and $E_8$ root systems these measures are generic \cite{Ito}).
We come thus to a natural question on whether one can give a similar meaning to
general elliptic hypergeometric integrals describing ${\mathcal N}=4$
SCIs and construct corresponding biorthogonal functions.
The first example of such biorthogonal functions in the univariate case
has been found in \cite{S2} and a particular $SP(2N)$-group
multivariable generalization of them has been constructed in \cite{Rains}.
For the exceptional root systems $\mathcal{N}=4$ SCIs define
the only currently known integrals pretending to such a role.

{\bf Acknowledgments.} We would like to thank I. A. Bandos, F. A. H.
Dolan,  Z. Komargodski, I. V. Melnikov, Y. Nakayama, V. Niarchos, A. F. Oskin, S. Theisen, and B. Wurm for
valuable discussions. We are indebted also to S. S. Razamat and Yu.
Tachikawa for useful remarks to the paper. The first author was
partially supported by RFBR grant no. 09-01-00271 and
the Heisenberg-Landau program.

\appendix
\section{Characters of the adjoint representations}

Here we list characters of the adjoint representations for simple Lie groups
$G$ depending on the maximal torus variables $z_j$.

For $SU(N)$ group one has $N$ variables $z_j$, $\prod_{j=1}^N z_j = 1,$ and
$$
\chi_{SU(N),adj}(z_1,\ldots,z_N) = \sum_{1 \leq i<j
\leq N} (z_iz_j^{-1} + z_i^{-1}z_j) +N-1.
$$

For $SO(2N+1)$ group of rank $N$ the
character is (no constraints on $z_j$)
$$
\chi_{SO(2N+1),adj}(z) = \sum_{1 \leq i < j \leq N} z_i^{\pm1}z_j^{\pm1}
+\sum_{i=1}^N z_i^{\pm1}+ N,
$$
where $z_i^{\pm1}z_j^{\pm1}:= z_iz_j + z_iz_j^{-1} + z_i^{-1}z_j +
z_i^{-1}z_j^{-1}$ and $z_i^{\pm1}:=z_i + z_i^{-1}$.

For $SP(2N)$ and $SO(2N)$ groups of rank  $N$ the characters are
\begin{eqnarray*} &&
\chi_{SP(2N),adj}(z) = \sum_{1 \leq
i < j \leq N} z_i^{\pm1}z_j^{\pm1} + \sum_{i=1}^Nz_i^{\pm2} + N,
\\ &&
\chi_{SO(2N),adj}(z) = \sum_{1
\leq i < j \leq N} z_i^{\pm1}z_j^{\pm1} + N.
\end{eqnarray*}

The character for the adjoint representation of $G_2$ group is a
symmetric polynomial of two parameters $z_1$ and $z_2$, but it is
convenient to introduce the third variable using relation
$z_1z_2z_3=1$. Then,
$$
\chi_{G_2,adj}(z_1,z_2,z_3) = 2 + \sum_{1 \leq i < j \leq 3} z_i^{\pm1}z_j^{\pm1}.
$$

The exceptional $F_4$ group has rank four and
\begin{eqnarray*}
&& \chi_{F_4,adj}(z_1,\ldots,z_4)
=\sum_{i = 1}^4 z_i^{\pm1}+ \sum_{1 \leq i < j \leq 4}
z_i^{\pm1}z_j^{\pm1}\nonumber \\
&& \makebox[2em]{} + (z_1^{1/2} + z_1^{-1/2})(z_2^{1/2} +
z_2^{-1/2})(z_3^{1/2} + z_3^{-1/2})(z_4^{1/2} + z_4^{-1/2}) +4.
\end{eqnarray*}

Description of the exceptional Lie groups $E_{6,7,8}$
can be found in \cite{Adams}. The rank of the group $E_6$ is equal to six and
\begin{eqnarray*} &&
\chi_{E_6,adj}(z_1,\ldots,z_6) = 6 + \sum_{1 \leq i < j \leq 5}
z_i^{\pm1}z_j^{\pm1} \nonumber \\
&& \makebox[-2em]{} + z_6^{3/2}\prod_{i=1}^5 z_i^{-1/2}
\Big(1 + \sum_{1 \leq i  < j  \leq 5} z_{i }
z_{j } + \sum_{1 \leq i  < j  < k  < l  \leq 5} z_{i } z_{j }
z_{k } z_{l } \Big) \\ \nonumber && \makebox[-2em]{} +
z_6^{-3/2}\prod_{i=1}^5 z_i^{1/2}\Big(1 + \sum_{1 \leq
i  < j  \leq 5} (z_{i } z_{j })^{-1} + \sum_{1 \leq i  < j  <
k  < l  \leq 5} (z_{i } z_{j } z_{k } z_{l })^{-1}\Big).
\end{eqnarray*}

The rank of the group $E_7$ is equal to seven and the needed character is
\begin{eqnarray*} &&  \makebox[-2em]{}
\chi_{E_7,adj}(z_1,\ldots,z_7) = 7 + \sum_{1
\leq i < j \leq 6} z_i^{\pm1}z_j^{\pm1} + z_7^{\pm2}
 \\ && \makebox[-2em]{} + (z_7
+ z_7^{-1})\Big(\prod_{l=1}^6 z_l^{1/2} \sum_{i = 1}^6
z_i^{-1} + \prod_{l=1}^6 z_l^{-1/2} \Big(\sum_{i =
1}^6 z_i + \sum_{1 \leq i < j < k \leq 6} z_i z_j z_k\Big)\Big).
\end{eqnarray*}

The group $E_8$ is the biggest exceptional Lie group, it has rank eight and
\begin{eqnarray*} &&
\chi_{E_8,adj}(z_1,\ldots,z_8) = 8 + \sum_{1 \leq i < j
\leq 8} z_i^{\pm1}z_j^{\pm1} + \prod_{i=1}^8 z_i^{-1/2} \Big(1 +
\sum_{1 \leq i  < j  \leq 8} z_{i } z_{j }\Big)
\nonumber \\
&& \makebox[2em]{} + \prod_{i=1}^8 z_i^{1/2} \Big(1 + \sum_{1 \leq
i  < j  \leq 8} (z_{i } z_{j })^{-1} + \sum_{1 \leq i  < j  <
k  < l  \leq 8} (z_{i } z_{j } z_{k } z_{l })^{-1}\Big).
\end{eqnarray*}

\end{document}